\LoadClass{birkmult}
\documentclass{birkmult}
\usepackage{graphicx}
\usepackage{amsmath}
\usepackage{mathrsfs}
\usepackage{upgreek}
                                                                                                                                                                                                  
\begin{document}

\title{Exact Description of Rotational Waves in an Elastic Solid  \footnote{Accepted 2010 by Advances in Applied Clifford Algebras }}
\author{R. A. Close}
%\date{}
\address{%
4110 SE Hawthorne Blvd. \#232 \\
Portland, OR 97214 \\
USA}
\email{robert.close@classicalmatter.org}

\begin{abstract} Conventional descriptions of transverse waves in 
an elastic solid are limited by an assumption of infinitesimally 
small gradients of rotation. By assuming a linear response to variations 
in orientation, we derive an exact description of a restricted 
class of rotational waves in an ideal isotropic elastic solid. 
The result is a nonlinear equation expressed in terms of Dirac 
bispinors. This result provides a simple classical interpretation 
of relativistic quantum mechanical dynamics. 
We construct a Lagrangian of the 
form $ \mathscr{L} =- \mathscr{E}+U+K=0 $, where $ \mathscr{E}$ is the total energy, \textit{U} is the potential energy, and \textit{K} 
is the kinetic energy. 
\end{abstract}

%PACS Codes: 03.65.Pm, 03.65.Ta, 11.10.Ef, 11.15.Kc, 46.05.+b, 
%46.15.Cc

\maketitle

\section{Introduction}

The ideal elastic solid has been an important model in the history 
of physics. It is a good approximation for many processes in 
condensed matter. It was also the basis of early theories of light 
because of its ability to support transverse waves \cite{Whittaker}. 
Aether models fell into disuse at the end of the 19$^{th}$ 
century when experiments 
failed to detect variations in light speed relative to the direction 
of the earth's motion through space \cite{Michelson}. However, de Broglie's  wave 
hypothesis for matter \cite{deBroglie} predicts a null result for such aether-drift 
experiments whether or not the waves propagate through a material medium.

It was recently proposed that a lattice of elastic cells would yield fermions 
and gauge fields equivalent to the Standard Model of particle 
physics \cite{Schmelzer}. This mechanical model is similar to 
the one James Maxwell used to develop the equations of electromagnetism 
\cite{Whittaker}. While a homogeneous elastic solid may 
be regarded as a continuum limit of such a lattice, it is fundamentally 
simpler in structure. 

It is current practice to describe physical events in terms of discrete elementary 
particles rather than continuous waves. However, the Dirac and 
Klein-Gordon equations which are used to compute the correlations 
between particle events also determine the wave-like evolution 
of continuously distributed physical quantities such as momentum, 
energy, and angular momentum densities \cite{Takabayashi, Hestenes}.
Furthermore, the first-order Dirac equation may be regarded 
as a representation of a second-order wave equation with the 
bispinor wave function encoding the first derivatives \cite{Matsutani, Rowlands, Cullerne, Rowlands2, Close}. Half-integer spin is attributable to the fact that waves propagating in opposite 
directions are independent states separated by 180-degree rotation. Thus 
it is quite possible that a single wave equation may determine 
the evolution of dynamical quantities everywhere in space.

An ideal homogeneous, isotropic, elastic solid (with no dissipation 
and constant elasticity) is the simplest possible model for wave-like processes 
in three spatial dimensions. Hence the physics of an ideal elastic 
solid should be the foundation for any analysis of Lorentz-covariant 
processes in three-dimensional space. Yet this simple model has 
never been properly understood. 

Prior analyses of the elastic 
solid rely on approximations of infinitesimally small rotations 
or small rates of variation of rotations \cite{Kleinert}. Analysis of first-order stress and strain yields an elastic wave equation for displacement ${\bf a} \left( {\bf r},t \right)$:
\begin{equation}
\rho \partial _{t} ^{2} {\bf a} = \left( 2 \mu + \lambda \right) \nabla \left[ \nabla \cdot {\bf a} \right]
- \mu \nabla \times \left[ \nabla \times {\bf a} \right]
\end{equation}

This equation has two obvious flaws. First, since the displacement field is defined with respect to positions in space rather than particular elements of the solid, the value of the field is transported with the solid as it moves. Inclusion of convection by velocity $\bf u$ and rotation by vorticity $\bf w$ yields:\begin{equation}
\rho \partial _{t} ^{2} {\bf a} +\rho {\bf u} \cdot \nabla \partial_t {\bf a} - \rho {\bf w} \times \partial_t {\bf a} = \left( 2 \mu + \lambda \right) \nabla \left[ \nabla \cdot {\bf a} \right]
- \mu \nabla \times \left[ \nabla \times {\bf a} \right]
\end{equation}

The second flaw is more problematic. A rigid rotation by angle $\boldsymbol \Theta$ yields 
a non-zero divergence of the displacement field ($\nabla \cdot {\bf a}$ $=2 \left( \cos {\it \Theta} -1\right) $). If $\boldsymbol \Theta$  varies  with position either radially or along the rotation  axis, then the motion  preserves  volume and the elastic response should be proportional to the shear modulus $\mu$. 
Instead, the term $ \left( 2 \mu + \lambda \right)  \nabla \left[ \nabla \cdot {\bf a} \right]$  fails to distinguish between compressional and rotational origins of divergence of the displacement field. 
This is why the equation is only valid for infinitesimally small gradients of rotation.

The goal of this paper is to formulate an exact 
description of rotational (or transverse) waves in an ideal elastic solid. We will accomplish this task by describing incompressible motion in terms of rotations rather than displacements.

\section{Basic Assumptions}

We make the following basic assumptions: \\
1. 
 The elastic solid is characterized by an inertial density $ \rho $ 
and coefficient of elasticity $\mu$, with characteristic wave speed $c=\sqrt{\mu /\rho }$.\\
2. 
There is a linear response to variations of orientation angle $\boldsymbol \Theta$
 relative to equilibrium. This means that an initial static perturbation 
(with velocity ${{\bf u} \left( {\bf r} \right)} = 0$) would yield the response:
\begin{equation}\label{linear_response}
\partial _{t}^{2} \boldsymbol \Theta=c^{2} \nabla ^{2} \boldsymbol \Theta
\end{equation}
3. 
The velocity field \textbf{u} has no divergence 
($\nabla \cdot {{\bf u}}=$0). Therefore the velocity may be written as the curl of a vector 
field:
\begin{equation}
{{\bf u}}=\frac{1}{2\rho } \left[ \nabla \times \bar{\bf J} \right] 
\end{equation}
The vector field $\bar{\bf J}$ is called the conjugate angular momentum density. It differs 
from the usual definition of angular momentum density ${\bf J}={\bf r}\times \rho {{\bf u}}$ in that it is independent of the choice of origin and can have arbitrary direction. If ${\bf u}$ falls to zero sufficiently rapidly toward infinity, then integration by parts yields the alternative expressions for kinetic energy:
\begin{equation}
 K=\frac{1}{2} \int d^{3}r \,\rho u^{2}  = \frac{1}{2} \int d^{3}r  \,{{\bf w}} \cdot \bar{\bf J} 
 \end{equation}
 where ${{\bf w}}=\nabla \times {{\bf u}}/2 $ is the angular velocity, or vorticity. Hence $\bar{\bf J} $
is the variable conjugate to angular velocity for a Lagrangian 
which depends on ${\bf u}$ only through the (positive) kinetic energy. \\
4.
We assume that there are no velocity-dependent forces (such as frictional damping). Hence velocity only enters the equations of evolution through the convection and rotation of fields. We neglect any effect of the symmetric tensor ($\partial_i u_j +\partial_j u_i$) on the orientation angle or angular momentum direction.

Additional assumptions will be introduced in order to simplify 
the mathematics, and these may limit the generality of the results.

\section{Equation of Evolution}

Starting from \eqref{linear_response}, we define an angular potential ${\bf Q}$ such that:
$\nabla ^{2} {{\bf Q}} =-4\rho \boldsymbol \Theta$. The static condition for ${\bf Q}$ is:
\begin{equation}
\nabla ^{2} \left\{  \partial _{t}^{2} {{\bf Q}}-c^{2} \nabla ^{2} {{\bf Q}} \right\} =0 \quad \left(\text{ if } {{\bf u}} \left({\bf r} \right) =0 \text{ everywhere} \right)
\end{equation}
Define the spin angular momentum as:\begin{equation}
{{\bf S}} \equiv \partial _{t} {{\bf Q}}
\end{equation}
When motion is present, it contributes to the time derivative $\partial _{t} {\bf S}$
only through convection ($- {\bf u}\cdot \nabla {\bf S}$) and rotation (${\bf w}\times {\bf S}$):
\begin{equation}
\nabla ^{2} \left\{ \partial _{t}^{2} {\bf Q}-c^{2} \nabla ^{2} {\bf Q}+{\bf u} \cdot \nabla
{\bf S} - {\bf w}\times {\bf S} \right\} =0
\end{equation}
From here on, we will consider only wave-like 
solutions satisfying:
\begin{equation}\label{waves}
\partial _{t}^{2} {\bf Q}-c^{2} \nabla ^{2} {\bf Q}+{\bf u} \cdot \nabla {\bf S} -{\bf w} \times {\bf S}=0
\end{equation}
For oscillatory solutions to this equation, the first two terms 
are always in phase ($\partial _{t}^{2} {\bf Q}-c^{2} \nabla ^{2} {\bf Q}$), whereas the velocity-dependent terms may have different phase. However, if the velocity-dependent terms do not add to zero then they must have the same temporal phase as the linear terms: 

\begin{equation}
{\bf u}\cdot \nabla {\bf S}- {\bf w}\times {\bf S}=  \Omega ^{2} \left( {\bf r} \right) {\bf Q}
\end{equation}
where $\Omega ^{2} \left( {\bf r} \right) $ is some function of position (more generally, 
$\Omega ^{2} \left( {\bf r} \right) $ could have different values for each component of $\bf Q$). Substitution yields:
\begin{equation}
\partial _{t}^{2} {\bf Q}-c^{2} \nabla ^{2} {\bf Q}+\Omega ^{2} \left( {\bf r} \right)  {\bf Q}=0
\end{equation}
If $\Omega ^{2} \left( {\bf r} \right) $ is constant and positive, then this is the Klein-Gordon equation.

Now our only remaining task is to solve for the velocity in terms 
of other wave variables. To do this we note that the above equation \eqref{waves} can 
be written in terms of a four-component complex Dirac bispinor ($\psi $) using the following identifications: 
\begin{align}
S_{j} \equiv \partial _{t} Q_{j}  &\equiv \frac{1}{2}  \left[ \psi ^{\dagger}
\sigma _{j} \psi \right] ^{}  \notag  \\ 
c\left[ \nabla \cdot {\bf Q} \right] &\equiv -\frac{1}{2}
 \left[ \psi ^{\dagger} \gamma ^{5} \psi \right]  \notag  \\
c^{2} \left\{ \nabla \times \nabla \times {\bf Q} \right\} _{j} &\equiv
-\frac{\text i}{2} c\epsilon _{ijk} \left\{ \left[ \partial _{i} \psi ^{\dagger} \right] \gamma
^{5} \sigma _{k} \psi -\psi ^{\dagger} \gamma ^{5} \sigma _{k} \partial _{i} \psi
\right\}  
\end{align}
This identification between bispinors and classical variables was made previously  \cite{Close2009}, where it was also assumed that the velocity could be derived from the spin. In this paper we will derive an expression for velocity from the conjugate momentum.

The matrices $\sigma _{j} $ are the Dirac spin matrices: 
\begin{align}
\sigma _{1} &=\left( 
\begin{array}{cccc}
0 & 1 & 0 & 0 \\
1 & 0 & 0 & 0 \\
0 & 0 & 0 & 1 \\
0 & 0 & 1 & 0
\end{array}
\right) ,\quad \sigma _{2} =\left( 
\begin{array}{cccc}
0 & -{\text i}  & 0 & 0 \\
{\text i} & 0 & 0 & 0 \\
0 & 0 & 0 & -{\text i} \\
0 & 0 & {\text i} & 0
\end{array}
\right) ,\notag \\
 \sigma _{3} &=\left( 
\begin{array}{cccc}
1 & 0 & 0 & 0 \\
0 & -1 & 0 & 0 \\
0 & 0 & 1 & 0 \\
0 & 0 & 0 & -1
\end{array}
\right) 
\end{align}
The matrices $c\gamma ^{5} \sigma _{j} $ are the Dirac velocity matrices. One representation for 
$\gamma ^{5} $ is:
\begin{equation}
\gamma ^{5} =\left( 
\begin{array}{cccc}
0 & 0 & 1 & 0 \\
0 & 0 & 0 & 1 \\
1 & 0 & 0 & 0 \\
0 & 1 & 0 & 0
\end{array}
\right) 
\end{equation}
The above identifications provide seven constraints on the eight free 
parameters of the Dirac bispinor. In terms of bispinors, the rotational wave 
equation \eqref{waves} is:

\begin{align}
\frac{\partial }{\partial t} \left[ \psi ^{\dagger} \sigma _{j} \psi \right]
&+c\partial _{j} \left[ \psi ^{\dagger} \gamma ^{5} \psi \right] -{\text i} c\epsilon
_{ijk} \left\{ \partial _{i} \psi ^{\dagger} \gamma ^{5} \sigma _{k} \psi -\psi
^{\dagger} \gamma ^{5} \sigma _{k} \partial _{i} \psi \right\}  \notag \\
&+{\bf u} \cdot \nabla \left[ \psi ^{\dagger} \sigma _{j} \psi \right] -\epsilon _{kij}
w_{k} \left[ \psi ^{\dagger} \sigma _{i} \psi \right] \,=0 
\end{align}
Expanding the derivatives yields:\begin{equation}
\psi ^{\dagger} \sigma _{j} \left[ \partial _{t} \psi +c\gamma ^{5} \boldsymbol{\upsigma} \cdot \nabla
\psi +{\bf u} \cdot \nabla \psi +{\bf w} \cdot \frac{{\text i}\boldsymbol{\upsigma} }{2} \psi \right] +\text{c.c.}=0
\end{equation}
where (c.c.) represents the complex conjugate. The Hermitian conjugate 
wave function may be regarded as an independent variable (the 
independent real and imaginary parts of the wave function are 
linear combinations of elements of $\psi $ and  $\psi ^{\dagger} $). Validity for arbitrary 
$\psi ^{\dagger} $ requires the terms in brackets to sum to zero. This yields the Dirac equation:\begin{equation}
\partial _{t} \psi +c\gamma ^{5} \boldsymbol{\upsigma} \cdot \nabla \psi +{\bf u} \cdot \nabla \psi
+{\text i} {\bf w}\cdot \frac {\boldsymbol{\upsigma}}{2} \psi +{\text i} \chi \psi =0
\end{equation}
where $\chi $ may be any operator with the property:\begin{equation}
 Re \left( \psi ^{\dagger} \sigma _{j} {\text i} \chi \psi \right) =0
\end{equation}
Since $\chi $ has no effect on the original equation for $\bf Q$, we assume it to be zero.

Now we construct a Lagrange density $ \mathscr{L}$. Lagrange's equations of motion 
for a field variable $\psi$  are:
\begin{equation}\label{euler}
\partial _{t} \frac{\partial  \mathscr{L}}{\partial \left[ \partial _{t} \psi
\right] } +\sum\limits_{j}\partial _{j} \frac{\partial  \mathscr{L}}{\partial \left[
\partial _{j} \psi \right]  }  -\frac{\partial  \mathscr{L}}{\partial \psi }
=0
\end{equation}
A similar equation holds with $\psi ^{\dagger} $ replacing $\psi$. A possible Lagrange density for rotational waves is therefore:\begin{equation}
 \mathscr{L}= -{\text i} \psi ^{\dagger} \partial _{t} \psi +\psi ^{\dagger}
\left[ -{\text i}c\gamma ^{5} \boldsymbol{\upsigma} \cdot \nabla \right] \psi +\psi ^{\dagger} \left[
-{\text i}  {\bf u} \cdot \nabla +{\bf w} \cdot \frac{\boldsymbol{\upsigma}}{2} \right] \psi
\end{equation}
Using the Hermitian conjugate of \eqref{euler}  yields simply 
$\partial  \mathscr{L}/\partial \psi  ^{\dagger} =0$. Since the complex conjuate of the equation of evolution is also valid, we can require the Lagrange density to be real:\begin{equation}
 \mathscr{L}={\text Re} \left\{ -{\text i} \psi ^{\dagger} \partial _{t} \psi +\psi ^{\dagger}
\left[ -{\text i}c\gamma ^{5} \boldsymbol{\upsigma} \cdot \nabla \right] \psi +\psi ^{\dagger} \left[
-{\text i}  {\bf u} \cdot \nabla +{\bf w} \cdot \frac{\boldsymbol{\upsigma}}{2} \right] \psi \right\}
\end{equation}
The conjugate momentum to the field $\psi $
is $p_{\psi } $:
\begin{equation}\label{field_momentum}
p_{\psi } =\frac{\partial  \mathscr{L}}{\partial \left[ \partial _{t} \psi \right] }
=-{\text i} \psi ^{\dagger} 
\end{equation}
We assume that we can neglect boundary terms in the integration 
by parts of:\begin{equation}
\int d^{3}r \, {\bf w} \cdot {\bf S} = \frac{1}{2} \int d^{3}r \, \left[ \nabla \times {\bf u} \right]
\cdot {\bf S} = \frac{1}{2} \int d^{3}r \, {\bf u} \cdot \left[ \nabla \times {\bf S} \right]  
\end{equation}
The conjugate momentum for $\bf r$ is:
\begin{equation}\label{momentum}
{\bf p}_{\bf r} =\frac{\partial  \mathscr{L}}{\partial \bf u} ={\text Re} \left( -\psi ^{\dagger}
{\text i} \nabla \psi \right) +\frac{1}{2} \nabla \times \psi ^{\dagger} \frac{\boldsymbol{\upsigma}}{2} \psi 
=\rho {\bf u}
\end{equation}
This conjugate momentum was derived under the assumption that \textbf{u} 
is an independent variable. Identification of ${\bf p}_{r} $ with 
$\rho \bf u$ is justified by the lack of external forces, implying that $\rho \bf u$
enters the Lagrangian only through the kinetic energy. 

Making $\bf u$ a function of $\psi $ introduces a factor of $1/2 $:
\begin{align}\label{Lagrange_final}
 \mathscr{L} &={\text Re} \left\{ -{\text i} \psi ^{\dagger} \partial _{t} \psi +\psi ^{\dagger}
\left[ -{\text i} c\gamma ^{5} \boldsymbol{\upsigma} \cdot \nabla \right] \,\psi \right\} \notag \\
&+\frac{1}{2\rho } \left[ {\text Re} \left( -\psi ^{\dagger} {\text i} \nabla \psi
\right) +\frac{1}{2} \nabla \times \psi ^{\dagger} \frac{\boldsymbol{\upsigma}}{2} \psi \right] ^{2} 
\end{align}
Variation of this Lagrange density determines the evolution of 
rotational waves: 
\begin{equation}
\partial _{t} \psi +c\gamma ^{5} \boldsymbol{\upsigma} \cdot \nabla \,\psi +{\bf u} \cdot \nabla \psi
+{\text i} {\bf w} \cdot \frac{\boldsymbol{\upsigma}}{2} \psi =0
\end{equation}
where velocity $\bf u$ and vorticity ${\bf w}=\nabla \times {\bf u}/2 $ are functions of $\psi$ as determined from \eqref{momentum}.

\section{Dynamical Variables}

\subsection{Angular Momentum}

Recall that the velocity is the curl of an origin-independent angular momentum:
\begin{equation}
{\bf u}=\frac{1}{2\rho } \nabla \times \bar{\bf J} 
\end{equation}
We can separate the conjugate angular momentum ($\bar{\bf J}$) into orbital ($\bar{\bf{L}}$) and spin ($\bf S$) components by identifying the orbital (${\bf u}_{\bf L} $) and spin (${\bf u}_{\bf S} $) contributions to velocity:

\begin{align}
{\bf u}_{\bf L} &=\frac{1}{\rho} {\text Re} \left( -\psi ^{\dagger}
{\text i} \nabla \psi \right) = \frac{1}{2 \rho} \nabla \times \bar{\bf L} \notag \\
{\bf u}_{\bf S} &=\frac{1}{2\rho} \nabla \times \psi ^{\dagger} \frac{\boldsymbol{\upsigma}}{2} \psi = \frac{1}{2 \rho} \nabla \times {\bf S} 
\end{align}
The incompressibility condition $\nabla \cdot {\bf u}=0$
 places an additional restriction on the wave function:
\begin{equation}
\nabla \cdot \rho {\bf u}_{\bf L} =\frac{1}{2} \nabla \cdot \left[ -\psi ^{\dagger} {\text i}
\nabla \psi +{\text i} \left[ \nabla \psi \right] ^{} \psi \right]
=\frac{{\text i} }{2} \left\{ -\psi ^{\dagger} \nabla ^{2} \psi +\left[ \nabla ^{2}
\psi ^{\dagger} \right] ^{} \psi \right\} =0
\end{equation}
The usual definition of angular momentum ${\bf J}={\bf L}+{\bf S}$ depends on the choice 
of origin. If we define a rotational velocity as 
${\bf u}_{\text R} ={\bf w} \times {\bf r}$
  and note that the vorticity is the instantaneous angular velocity 
(${\bf w}=d{\boldsymbol \Theta}/dt$), then from (19) the conjugate angular momentum would have the same form as found in relativistic quantum mechanics:
\begin{equation}
{\bf p}_{\boldsymbol \Theta} =\frac{\partial {\mathscr L}}{\partial \left[ \partial _{t} {\boldsymbol \Theta} \right] } =\psi ^{\dagger}
\left\{ -{\bf r}\times {\text i} \nabla +\frac{\boldsymbol{\upsigma}}{2} \right\} \psi ={\bf L}+{\bf S}
\end{equation}

\subsection{Energy and momentum}

In the Lagrange density defined in \eqref{Lagrange_final}, the velocity-dependent 
term is clearly the kinetic energy density. This observation 
suggests that the Lagrangian has the form:\begin{equation}
{\mathscr L}={\text Re} \left\{ -{\text i} \psi ^{\dagger} \partial _{t} \psi +\psi ^{\dagger}
\left[ -{\text i} c\gamma ^{5} \boldsymbol{\upsigma}  \cdot \nabla \right] \,\psi +\frac{1}{2} \rho
u^{2} \right\} =-{\mathscr E}+U+K
\end{equation}
where ${\mathscr E} \equiv {\text Re} \left( {\text i} \psi ^{\dagger} \partial _{t} \psi \right) $
 is the total energy, 
$U \equiv {\text Re} \left( -\psi ^{\dagger} \left[ {\text i} c\gamma ^{5} \boldsymbol{\upsigma} \cdot \nabla
\right] \,\psi \right) $
 is potential energy, and \textit{K} is kinetic energy density. We will simply adopt these definitions without further justification.

The Hamiltonian is the negative of the energy:
\begin{align}
{\mathscr H} =p_{\psi } \partial _{t} \psi -{\mathscr L}
&={\text Re} \left\{ \psi ^{\dagger} \left[ {\text i} c\gamma ^{5} \boldsymbol{\upsigma}
\cdot \nabla +\frac{1}{2} \left( {\text i} {\bf u} \cdot \nabla -{\bf w} \cdot \frac{\boldsymbol{\upsigma}}{2}\right) \right] \psi \right\}  \notag \\
&=-\left\{ U+\frac{1}{2} \rho u^{2} \right\} 
\end{align}
Hamilton's equation for the wave function is:\begin{equation}
\partial _{t} \psi =\frac{\partial {\mathscr H}}{\partial p_{\psi } }
=\frac{\partial {\mathscr H}}{\partial \left[ -{\text i} \psi ^{\dagger} \right] ^{^{} } }
=\left\{ -c\gamma ^{5} \boldsymbol{\upsigma} \cdot \nabla -{\bf u}\cdot \nabla -{\text i} {\bf w}\cdot
\frac{\boldsymbol{\upsigma}}{2} \right\} \psi 
\end{equation}
We can also define a Hamiltonian operator with 
$\partial _{t} \psi ={\text i} H\psi $
(note opposite sign convention from quantum mechanics):\begin{equation}
H={\text i} c\gamma ^{5} \boldsymbol{\upsigma} \cdot \nabla +{\text i} {\bf u} \cdot \nabla -{\bf w}\cdot
\frac{\boldsymbol{\upsigma}}{2} 
\end{equation}

The Hamiltonian is a special case ($T_{0}^{0} $) of the energy-momentum tensor:
\begin{equation}
T_{\nu }^{\mu } =\frac{\partial {\mathscr L}}{\partial \left[ \partial _{\mu } \psi
\right] ^{^{} } } \partial _{\nu } \psi -{\mathscr L} \delta _{\nu }^{\mu } 
\end{equation}
The dynamical momentum density $\bf P$ is derived from the orbital part 
of the angular momentum:
\begin{equation}
T_{i}^{0} =P_{i} =-{\text i} \psi ^{\dagger} \partial _{i} \psi 
\end{equation}
This is identical to the dynamical momentum of relativistic quantum mechanics. Unlike the conjugate momentum, it does not include the rotational motion associated with the spin angular momentum.

The sign of the Hamiltonian (and Lagrangian) is simply a convention. We chose signs to make the conjugate momentum parallel (rather than anti-parallel) to velocity, thus requiring the Hamiltonian to be the negative of the energy.
In analogy with plane waves, the function 
$\cos \left( \omega t-kx\right) $ with $\omega >0$
 is equivalent to $\cos \left( \omega t+kx\right) $
 with $\omega <0$. We could change the sign of the Lagrangian and Hamiltonian 
and still preserve the sign of the momentum by using covariant 
derivatives ($\partial _{\mu } =\left( \partial _{t} ,-\partial _{i} \right) $). 
However, such a relativistic construction is unnecessary and 
perhaps misleading since we are in fact dealing with Galilean 
space-time (the laws of relativity apply to the space of measurements, not to the absolute space-time). The energy-momentum tensor components
$T_{\mu }^{0} =\left( -{\mathscr E},P_{i} \right) $
 may still be regarded as a covariant vector whose magnitude 
is the scalar ${\mathscr E}^{2} -P^{2} $. 
The equation of evolution is of course unaffected by the choice 
of sign of the Lagrangian.

\section{Discussion}

The description of rotational waves derived here provides a simple 
classical interpretation of Dirac bispinors and quantum mechanical 
operators. Due to the nonlinear nature of the equation of evolution, 
we expect that an elastic solid has a discrete spectrum of fermionic 
soliton solutions. If each soliton is assigned its own independent 
wave function, then cancellation of interference between these 
independent `particles' would yield the Pauli exclusion principle 
and require the introduction of interaction potentials. \cite{Close2009} Characterization 
of these interaction potentials is beyond the scope of this paper.

Correlations between rotational soliton waves would be computed in the same manner as in relativistic quantum mechanics.\cite{Close2009} The reason is that the bispinor wave functions, and not the measurement values, satisfy a first order wave equation which may be used to directly compute variations in space and time. Therefore Bell's Theorem \cite{Bell} is not applicable.

The quantum mechanical Dirac equation is also equivalent to a 
deterministic equation for the evolution of spin angular momentum 
density. Comparison of the classical and quantum equations for 
spin density may yield new understanding of the physical properties 
of elementary particles. 

\section{Conclusions}

A nonlinear Dirac equation describes evolution of rotational 
waves in an ideal isotropic elastic solid. The equation is derived 
by assuming a linear response to variations in orientation and incorporating 
the influence of motion through convection and rotation. The Hamiltonian 
is represented as a sum of potential and kinetic energy. This 
result provides a simple classical interpretation of Dirac bispinors 
and the associated dynamical operators of relativistic quantum mechanics.

% ------------------------------------------------------------------------

\subsection*{Acknowledgment}
The author is grateful to Damon Merari for his interest and encouragement.

\end{document}